\documentclass[preprint,pre,showpacs,preprintnumbers]{revtex4}
\usepackage{amssymb}

\usepackage{graphicx}

\begin{document}

\title{Effects of anchored flexible polymers on mechanical properties of model biomembranes}

\author{Hao Wu} \email[Email address: ]{wade@issp.u-tokyo.ac.jp, wadewizard@gmail.com}
\affiliation{Institute for Solid State Physics, University of Tokyo,
Kashiwa 277-8581, Chiba, Japan \\ Department of Physics, School of
Science, University of Tokyo, 7-3-1 Hongo, Bunkyo-ku, Tokyo
113-0033, Japan}
\author{Hiroshi Noguchi}\email[Email address: ]{noguchi@issp.u-tokyo.ac.jp}
\affiliation{Institute for Solid State Physics, University of Tokyo,
Kashiwa 277-8581, Chiba, Japan \\ Department of Physics, School of
Science, University of Tokyo, 7-3-1 Hongo, Bunkyo-ku, Tokyo
113-0033, Japan}


\begin{abstract}
We have studied biomembranes with grafted polymer chains using a
coarse-grained membrane simulation, where a meshless membrane model
is combined with polymer chains. We focus on the polymer-induced
 entropic effects on mechanical properties
 of membranes. The spontaneous curvature
 and bending rigidity of the membranes increase with increasing polymer density.
Our simulation results agree with the previous theoretical
predictions.
\end{abstract}
\pacs{87.16.D-,87.17.Aa,82.70.Uv}

\keywords {spontaneous curvature, bending rigidity, anchored
polymer}

\maketitle

\section{INTRODUCTION}
Cellular membranes in living cells are a complex and compound
system, including various kinds of proteins and different lipid
compositions. Some of them are "decorated" by  sugar chains called
glycoproteins or glycolipids \cite{A:Alberts2007,A:LipSac1995}. From
theoretical and experimental aspects, pure lipid membrane in
tensionless fluidic state has extensively received much attention
for a long time \cite{A:Seifert1997}. Pioneer studies on the
polymer-membrane compound system, especially on membrane with
anchored polymers, have been recently reported by several groups
both theoretically and experimentally
\cite{A:Li1995,A:Li1996,A:Auth2003,A:HG1999}. They found that the
mechanical properties of  lipid membrane, such as the bending
rigidity $\kappa$, the saddle-splay modulus $\bar{\kappa}$, and the
spontaneous curvature $c_0$, can be dramatically changed by the
interaction between membrane and the surrounding solvent and
macromolecules \cite{A:LiPA1998,A:LiHG1998}. Liposomes decorated by
grafted polymers (PEG etc.) are often used for carriers in drug
delivery systems \cite{A:Needham1992,A:Allen2002}. Thus, it is
important to understand the effects of grafted polymers for both
fundamental research and medical applications.

In this paper, we explain our simulation model and then briefly
describe  the effects of grafted polymers on membrane properties. We
employed a meshless membrane model with grafted polymer chains,
since it allows  large scale simulations and the mechanical
properties of pure membranes are easy to control
\cite{A:nog09,A:Hiro2006}. For an isolated anchored polymer chain,
the membrane properties have been investigated in details by using
scaling analysis and partition function method
\cite{A:Li1995,A:Li1996,A:Auth2003}. Here we focus on membranes with
a high polymer density, where the interactions between polymer
chains are not negligible and have significant effects.

\section{model and method}

\begin{figure}
  \resizebox{18pc}{!}{\includegraphics{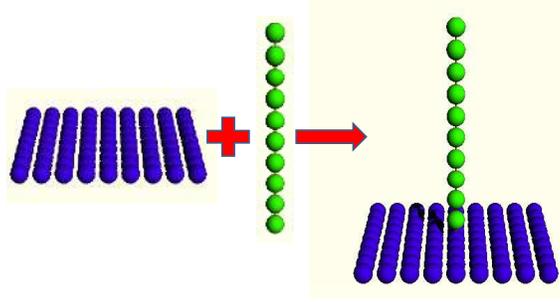}}
\caption{\label{fig:cart} A coarse-grained lipid membrane anchored
with a coarse-grained polymer chain to form a grafted biomembrane
system. The dark-gray particle represent membrane particles. A
single polymer chain is composed of ten (light-gray) particles which
are freely linked as an excluded-volume chain.}
\end{figure}

In this work, we employ a coarse-grained lipid membrane grafted with
a coarse-grained flexible linear polymer chain to form a grafted
biomembrane system (see Fig.~\ref{fig:cart}). One membrane particle
represents a patch of bilayer membrane, and possesses only
transnational degree of freedom. The particles form a quasi-2D
membrane by a curvature potential based on moving least-squares
(MLS) method \cite{A:Hiro2006}. (ii) Polymer particles are linked by
a harmonic potential, and freely move like a bead-spring behavior
with a soft-core repulsion.  An end of each polymer chain is
anchored on one membrane particle with the harmonic potential.

We consider a patch of membrane composed of $N_{\rm {mb}}$ membrane
particles. Among them, $N_{\rm {chain}}$ membrane particles are
anchored by a polymer chain. Each polymer chain consists of  $N_{\rm
{p}}$ polymer segments with a connected membrane particle. The
membrane and polymer particles interact with each other via a
potential
\begin{equation}
U_{\rm {tot}}=  U_{\rm {rep}} + U_{\rm {mb}} +  U_{\rm {p}}.
\label{totpot}
\end{equation}
All particles have a soft-core excluded volume term with a diameter
of $\sigma$.
\begin{equation}
U_{\rm
{rep}}=\varepsilon\sum_{i<j}\exp\left[-20(r_{ij}/\sigma-1)+B\right]f_{\rm
{cut}}(r_{ij}/\sigma) \label{rep}
\end{equation}
which $r_{ij}$ is the distance between membrane (or polymer)
particles $i$ and $j$. The diameter $\sigma$ is used as the length
unit, $B=0.126$, and $f_{\rm {cut}}(s)$ is a $C^{\infty}$ cutoff
function
\begin{equation}
f_{\rm {cut}}(s)=\left\{
\begin{array}{ll}
\exp\left\{A\left[1+\frac{1}{(|s|/s_{\rm {cut}})^n-1}\right]\right\}
&(s<s_{\rm {cut}})
 \\
0 &(s \geqslant  s_{\rm {cut}})
\end{array}
\label{cutoff} \right.
\end{equation}
with $n=12$. The factor $A$ in Eq.~$(3)$ is determined so that
$f_{\rm {cut}}(s_{\rm {half}})=0.5$, which implies $A=\ln(2){(s_{\rm
{cut}}/s_{\rm {half}})^n-1}$. In Eq.~$(2)$, we use the parameters
$A=1$, and $s_{\rm {cut}}=1.2$.

\subsection{Meshless membrane model}

The membrane potential $U_{\rm {mb}}$ consists of attractive and
curvature potentials,
\begin{equation}
U_{\rm {mb}}=\varepsilon \sum_i^{N_{\rm {mb}}} U_{\rm {att}}(\rho_i)
+k_{\alpha}\alpha_{\rm {pl}}({\mathbf{r}}_{i}), \label{mempotential}
\end{equation}
where the summation is taken only over the membrane particles. The
attractive multibody potential is employed to mimic the
"hydrophobic" interaction.
\begin{equation}
U_{\rm {att}}(\rho_i)=
0.25\ln\left\{1+\exp\left[-4(\rho_{i}-\rho^{\ast})\right]\right\}-C,
\label{att}
\end{equation}
which is a function of the local density of membrane particles
\begin{equation}
\rho_{i}=\sum_{j\neq i}^{N_{\rm {mb}}} f_{\rm {cut}}(r_{ij}/\sigma),
 \label{dens}
\end{equation}
with $n=12$, $s_{\rm {half}}=1.8$ [satisfied with $f_{\rm
{cut}}(s_{\rm {half}})=0.5$], and $s_{\rm {cut}}=s_{\rm
{half}}+0.3$. The constant
$C=0.25\ln\left[1+\exp(4\rho^\ast)\right]$ is chosen so that $U_{\rm
{att}}=0$ at $\rho_i=0$. Here we set $\rho^{\ast}=6$ in order to
simulate 2D fluidic membrane. For $\rho_{i}<\rho^{\ast}$, $U_{\rm
{att}}$ acts like a pairwise potential with $U_{\rm
{att}}=-2\sum_{j> i}f_{\rm {cut}}(r_{ij}/\sigma)$. For
$\rho_{i}\gtrsim \rho^{\ast}$, this potential saturates to the
constant $-C$. Thus, it is a pairwise potential with a cutoff at
higher densities than $\rho^\ast$.

The curvature potential is given by the
 shape parameter called "aplanarity", which is defined by
\begin{equation}
\alpha_{\rm {pl}}=\frac{9D_{\rm {w}}}{T_{\rm w}M_{\rm w}} ,
\label{apla}
\end{equation}
with the determinant $D_{\rm w}=\lambda_1\lambda_2\lambda_3$, the
trace $T_{\rm w}=\lambda_1+\lambda_2+\lambda_3$, and the sum of
principal minors $M_{\rm
w}=\lambda_1\lambda_2+\lambda_2\lambda_3+\lambda_3\lambda_1$. The
aplanarity $\alpha_{\rm {pl}}$ scales the degree of deviation from
the planar shape, and $\lambda_1$, $\lambda_2$, $\lambda_3$ are
three eigenvalues of the weighted gyration tensor
\begin{equation}
a_{\alpha\beta}({\mathbf{r}}_{i})=\sum_j^{N_{\rm {mb}}}
(\alpha_j-\alpha_G)(\beta_j-\beta_G)w_{\rm{cv}}(r_{ij}),
\label{eigen}
\end{equation}
where $\alpha,\beta\in \left\{x,y,z\right\}$. Without loss of
generality, we suppose $\lambda_1\leqslant \lambda_2\leqslant
\lambda_3$. If $\lambda_1$ is the minimum eigenvalue representing a
relative deviation from the local plane patch formed by the neighbor
membrane particles around, its corresponding eigenvector is
collinear with the normal vector $\mathbf{n}$ of this plane patch.
When the $i$-th membrane particle has two or less neighbor particles
within the cutoff distance $r_{\rm {cc}}$, they could be localized
on a certain plane, therefore $\alpha_{\rm {pl}}=0$. The mass center
of local plane patch
$\mathbf{r}_G=\sum_j\mathbf{r}_jw_{\rm{cv}}(r_{ij})/\sum_j
w_{\rm{cv}}(r_{ij})$, where a Gaussian $C^\infty$ cutoff function is
employed to calculate the weight of the gyration tensor
\begin{equation}
w_{\rm{cv}}(r_{ij})=\left\{
\begin{array}{ll}
\exp\left[\frac{(r_{ij}/r_{\rm {ga}})^2}{(r_{ij}/r_{\rm
{cc}})^n-1}\right] &(r_{ij}<r_{cc})
 \\
0 &(r_{ij}\geqslant r_{\rm {cc}}),
\end{array}
\label{cutoff} \right. \label{weight}
\end{equation}
which is smoothly cut off at $r_{ij}=r_{\rm {cc}}$. Here we use the
parameters $n=12$, $r_{\rm {ga}}=0.5r_{\rm {cc}}$, and $r_{\rm
{cc}}=3\sigma$. The bending rigidity and the line tension of
membrane edges are linearly dependent on $k_{\alpha}$ and
$\varepsilon$, respectively, for $k_{\alpha} \gtrsim 10$, so that
they can be independently varied by changing $k_{\alpha}$ and
$\varepsilon$, respectively. As described in the previous work
\cite{A:Hiro2006}, the aplanarity $\alpha_{\rm {pl}}\simeq
9\lambda_1/\left(\lambda_2+\lambda_3\right)$ for the smallest
eigenvalue $\lambda_1\ll \lambda_2$,$\lambda_3$. Locally a membrane
patch can be expressed for small fluctuations in the Monge
representation as
\begin{equation}
z = z_0+\frac{1}{2}C_1(x-x_0)^2 +\frac{1}{2}C_2(y-y_0)^2,
\label{eq:Mong}
\end{equation}
where $x$ and $y$ coordinates are set to the principal axes. $C_1$
and $C_2$ are the principal curvatures of the membrane patch. In the
case of cylindrical membrane, $C_1=1/R$ and $C_2=0$. By averaging
over the local neighborhood with a weight function $w(r)$ where
$r^2=x^2+y^2$, we have
\begin{equation}
\lambda_1 = \langle z^2\rangle-\langle z\rangle^2 = (
C_1^2+C_2^2)\langle (r^2-\langle r^2\rangle )^2\rangle .
\label{eq:eigen}
\end{equation}
Since $C_1^2+C_2^2=(C_1+C_2)^2-2C_1C_2$, we easily know
\begin{equation}
k_{\alpha}\alpha_{\rm {pl}}\sim k_{\alpha}\lambda_1\sim k_{\alpha}(
C_1+C_2)^2.\label{eq:apl}
\end{equation}
Thus, a linear relation between the bending energy in Helfrich's
macroscopic model and our mesoscopic parameter is obtained as
$\kappa \propto k_{\alpha}$, which is numerically confirmed in the
previous work \cite{A:Hiro2006,A:Hay2011}.

The membrane has zero spontaneous curvature without polymers. The
details of this membrane model are described in Ref.
\cite{A:Hiro2006}.

\subsection{Polymer Chain}

In each polymer chain, its polymer particles are connected by a
harmonic potential.
\begin{equation}
U_{\rm p}=k_{\rm {har}} \sum_{\rm chain} U_{\rm {har}}(r_{i,i+1}),
\label{mempolypoten}
\end{equation}
where the summation is taken only between neighboring particles in
each polymer chain and between the end polymer particles and
anchored membrane particles. The harmonic potential is given by
\begin{equation}
U_{\rm {har}}(r)=\frac{1}{2}(r-b)^2, \label{polyharmonic}
\end{equation}
which  $b$ is a Kuhn length of polymer chain. We choose
$b=1.2\sigma$ here such that a polymer chain stays in the force-free
state for $r_{i,i+1}=b$.

\subsection{Simulation method}

The $NVT$ ensemble (constant number of particles $N$, volume $V$,
and temperature $T$) is used with periodic boundary conditions in a
simulation box of dimensions $L_x\times L_y\times L_z$. The dynamics
of both membrane and anchored flexible polymers are calculated by
using underdamped Langevin dynamics. The motions of membrane and
polymer particles are  governed by
\begin{equation}
m\frac{d^2\mathbf{r}_{i}}{dt^2}=-\frac{\partial U_{\rm
{tot}}}{\partial
\mathbf{r}_{i}}-\zeta\frac{d\mathbf{r}_{i}}{dt}+\mathbf{g}_{i}(t)
\label{Lang1}
\end{equation}
where $m$ is the mass of a particle (membrane or polymer particle)
and $\zeta$ is the friction constant. $\mathbf{g}_{i}(t)$ is a
Gaussian white noise, which obeys the fluctuation-dissipation
theorem:
\begin{eqnarray}
\langle g_{i,\alpha}(t) \rangle & =& 0, \\
\langle g_{i,\alpha}(t) g_{j,\beta}(t')\rangle & =&  2 k_{\rm
B}T\zeta \delta _{ij}\delta _{\alpha\beta} \delta(t-t'),\nonumber
\end{eqnarray}
where $\alpha,\beta \in \{x,y,z\}$ and $k_{\rm B}T$ is the thermal
energy. We use $N_{\rm p}=10$, $\varepsilon=4$, $k_{\alpha}=10$, and
$k_{\rm {har}}=10$ through this work.

\section{results}

\begin{figure}
  \includegraphics[width=7.8cm]{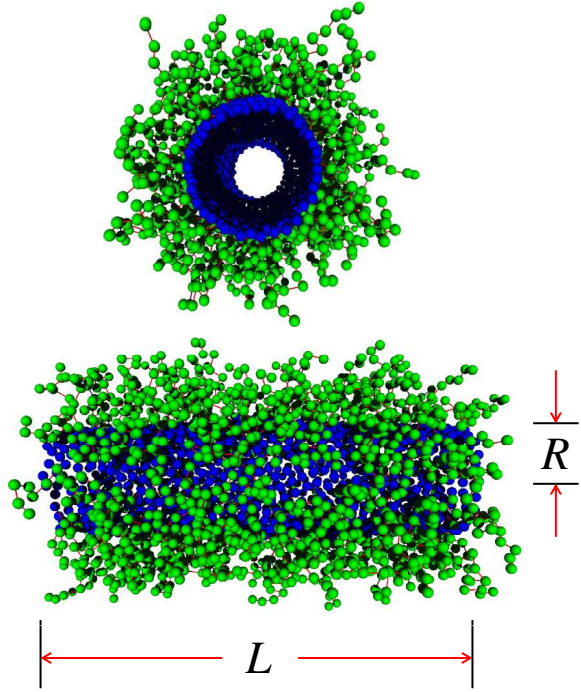}
  \caption{\label{fig:snap} Front and lateral snapshots of a cylindrical membrane with
anchored flexible polymers outside used for simulations. It contains
$1200$ membrane particles and $120$ polymer chains.}
\end{figure}

\begin{figure}
  \resizebox{18pc}{!}{\includegraphics{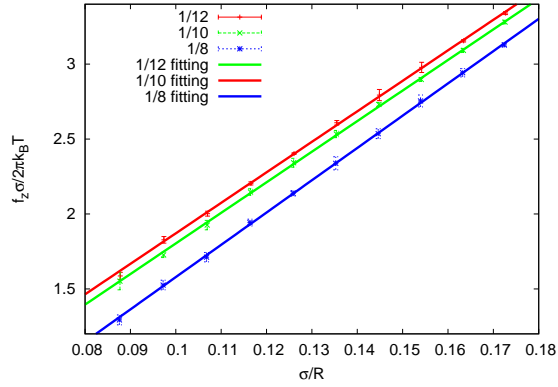}}
  \caption{\label{fig:fz}
Force $f_z$ dependence on the radius $R$ of the cylindrical
membranes with different polymer densities $\phi=1/12$, $1/10$, and
$1/8$, respectively. The solid lines are obtained by linear
least-squares fits.}
\end{figure}

A cylindrical membrane, composed of $N_{\rm {mb}}=1200$, with
$N_{\rm {chain}}$ anchored polymers outside (see
Fig.~\ref{fig:snap}) is used to estimate the polymer-induced
spontaneous curvature and bending rigidity. For a cylindrical
membrane with a radius $R$ and a length $L_z$, the curvature free
energy is given by
\begin{eqnarray}
F_{\rm {cv}} &=& \int \Big[ \frac{\kappa}{2} (C_1 + C_2- C_0)^2 +
\bar{\kappa} C_1C_2 \Big] dA \nonumber \\ \label{eq:F_tube}
 &=& 2\pi RL_z \left[ \frac{\kappa}{2}\left( \frac{1}{R} -C_0 \right)^2 \right],
\end{eqnarray}
where $C_1$ and $C_2$ are the principal curvatures at each position
on the membrane surface, and the membrane area $A=2\pi RL_z$. The
coefficients $\kappa$ and $\bar{\kappa}$ are the bending rigidity
and saddle-splay modulus, respectively. Biomembranes with lipids
symmetrically distributing in both leaflets of the bilayer have zero
spontaneous curvature $C_0=0$. The anchored polymer outside induces
asymmetry to the membrane.

The bending energy of the membrane generates a shrinking force along
the cylindrical axis,
\begin{equation}
f_z =\frac{\partial F_{\rm {cv}}}{\partial L_z} = 2\pi\kappa
\Big(\frac{1}{R} - C_0\Big). \label{eq:fz}
\end{equation}
Since a shorter cylinder membrane has a larger cylindrical radius,
it has a lower bending energy. Previously, we clarified that this
force can be used to estimate the spontaneous curvature $C_0$ and
bending rigidity $\kappa$ in simulations \cite{A:Hay2011}.
 We calculated $C_0$ and
$\kappa$ for another meshless model, which has an orientational
degree of freedom and can have a finite $C_0$. The obtained values
of $C_0$ agree very well with those calculated from an alternative
method using a curved membrane strip.

\begin{figure}
  \resizebox{18pc}{!}{\includegraphics{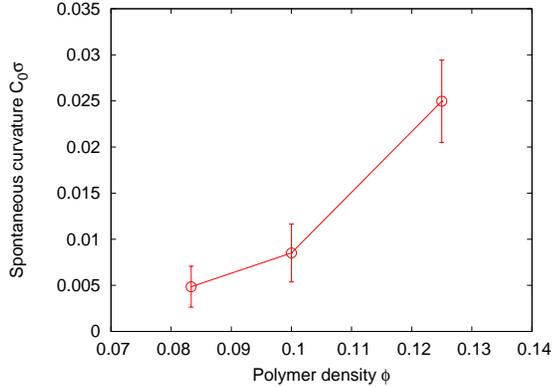}}
  \caption{\label{fig:spcv}
Polymer-induced spontaneous curvature is estimated by the linear
fitting method. It increases with increasing anchored polymer
density.}
\end{figure}

\begin{figure}
  \resizebox{18pc}{!}{\includegraphics{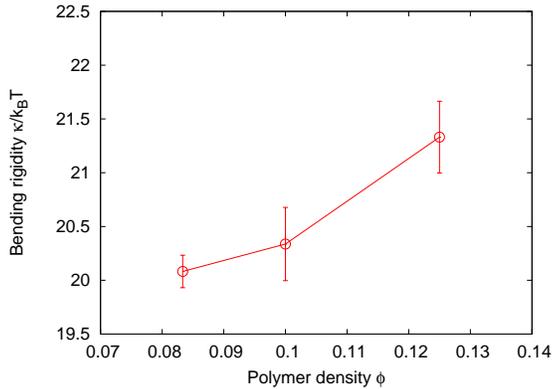}}
  \caption{\label{fig:bd}
Bending rigidity is estimated by the linear fitting method. It
increases with increasing anchored polymer density.}
\end{figure}

We define a polymer density $\phi=N_{\rm {chain}}/N_{\rm {mb}}$ to
quantitatively describe the relation between the change of
mechanical properties of membranes and the anchored polymer
concentration. Figures~\ref{fig:fz} shows the axial force $f_z$
calculated from the pressure tensor with different polymer
densities. The force $f_z$ increases linearly with $1/R$. Thus,
$C_0$ and $\kappa$ of the grafted membranes can be estimated from
the least squares fits to Eq. (\ref{eq:fz}). The obtained values of
$C_0$ and $\kappa$ are shown in Figs.~\ref{fig:spcv} and
\ref{fig:bd}, respectively. Both quantities increase with increasing
anchored polymer density. This quantitatively agrees with the
previous theory for a low polymer density. When the membrane is
curved in the opposite direction from the polymer grafting, the
polymer chains have more space to move. These entropic effects
 generate the spontaneous curvature of the membranes.
 The details of our results and comparison with the theory
will be described elsewhere~\cite{A:Hao2013}.

\section{summary}

We have studied the effects of grafted polymer on the membrane
properties. Anchored polymer chains are added to a meshless membrane
model. The free energy of polymer chains is dominated by the
entropic effects induced by the degree of freedom of polymer
conformation. We confirm that the induced spontaneous curvature and
bending rigidity both increase as the polymer chain density
increases. Anchored polymer chains also affect other properties of
biomembranes (it will be described in Ref. \cite{A:Hao2013}). The
properties and stability of membranes can be controlled by the
decoration of membrane with grafted polymers.

\section*{Acknowledgments}

We thank H. Shiba for helpful discussions. HW acknowledges the
support by a MEXT scholarship (Japan). This study is partially
supported by a Grant-in-Aid for Scientific Research on Priority Area
``Molecular Science of Fluctuations toward Biological Functions''
from the Ministry of Education, Culture, Sports, Science, and
Technology of Japan.

\end{document}